# Critical properties of the double exchange ferromagnet $Nd_{0.6}Pb_{0.4}MnO_3$

M. Sahana[1*], U. K. Rößler[1], Nilotpal Ghosh[2], Suja Elizabeth[2], H. L. Bhat[2], K. Dörr[1], D. Eckert[1], M. Wolf[1], K.–H. Müller[1]

[1] IFW Dresden, Postfach 270116, D-01171 Dresden, Germany.
[2] Department of Physics, Indian Institute of Science, Bangalore, India.

Results of a study of dc-magnetization $M(T, H)$, performed on a $Nd_{0.6}Pb_{0.4}MnO_3$ single crystal in the temperature range around $T_C$ (Curie temperature) which embraces the critical region $|\varepsilon| = |T - T_C|/T_C \leq 0.05$ are reported. The magnetic data analyzed in the critical region using the Kouvel-Fisher method give the values for the $T_C = 156.47 \pm 0.06$ K and the critical exponents, $\beta = 0.374 \pm 0.006$ (from the temperature dependence of magnetization), and $\gamma = 1.329 \pm 0.003$ (from the temperature dependence of initial susceptibility). The critical isotherm $M(T_C, H)$ gives $\delta = 4.547 \pm 0.1$. Thus the scaling law $\gamma + \beta = \delta\beta$ is fulfilled. The critical exponents obey the single scaling-equation of state $M(H, \varepsilon) = \varepsilon^\beta f_\pm(H/\varepsilon^{\beta+\gamma})$ where, $f_+$ for $T > T_C$ and $f_-$ for $T < T_C$. The exponent values are very close to those expected for the universality class of 3D Heisenberg ferromagnets with short-range interactions.

* Corresponding author
  Present address :   Kamerlingh Onnes Laboratory
                      Leiden Institute of Physics
                      Leiden University
                      P.O. Box 9504
                      2300 RA Leiden
                      The Netherlands
                      email: sahana@phys.leidenuniv.nl

*I. INTRODUCTION*

The relation between electrical transport and magnetism in colossal magnetoresistive manganites is one of the current challenges in solid-state physics[1]. The close interplay between transport and magnetism in these materials is widely attributed to the double-exchange (DE) mechanism[2], where Mn $e_g$ electrons hop between neighboring sites via oxygen $2p$ orbitals and align localized $t_{2g}$ spins due to strong intra atomic Hund's rule coupling. Correspondingly, at a finite-temperature paramagnetic-ferromagnetic (PM-FM) transition, the carriers in ferromagnetically aligned regions start to move freely through the lattice. The validity of this explanation of the colossal magnetoresistance (CMR) in doped ferromagnetic manganites has been questioned based on quantitative comparison with experiment[3]. The central unsolved problem concerns the importance and role of couplings of the conduction electrons that are responsible for the CMR and its apparent relation to the ferromagnetic transition to other modes, such as Jahn-Teller distortions, charge/orbital order, and phonons. It has been suggested that such couplings may result in various forms of polarons[4] in the manganites. Then, the PM-FM transition may be understood as a transition occurring when polaronic charge carriers become delocalized. Generalizing this picture, Archibald, Zhou, and Goodenough argued that the PM-FM transition should be first-order[5], triggered by a discontinuous transition between polaronic and extended itinerant states of the charge carriers. Both, for the basic DE-mechanism and for such extended models, the effective coupling between the localized core-like $t_{2g}$ spins is mediated by charge carriers subject to the fluctuations of the spins themselves and, therefore, may depend on the electron kinetics[6]. Other microscopic mechanisms for the ferromagnetism in doped manganites have been suggested based on the observation that the undoped reference state should be a charge-transfer insulator and thus, ferromagnetism in doped manganites should not originate from the usual double exchange interaction[7] (see also Ref. 1[d]). In our presentation and for definiteness, we follow the widely held opinion that a double exchange mechanism is responsible for ferromagnetism in this class of mixed-valent manganites. However, irrespective of the detailed nature of the microscopic couplings, the basic assumptions are existence of correlated and mobile charge-carriers responsible both for aligning the magnetic subsystem and for the transport. Within a ferromagnetically aligned region this may yield, e.g., long-range and/or multispin interactions. These general mechanisms are distinct from the localized-spin exchange interactions in a Heisenberg model. Thus, it is not clear how $Nd_{0.6}Pb_{0.4}MnO_3$ will behave at the critical point. Claiming the essential validity of the DE-model[8] Motome and Furukawa argue, based on computational studies for simplified DE-models, that the PM-FM transition in the CMR manganites should belong to the short-range Heisenberg universality class[6,9]. Limitations of these computational studies do not allow to determine precisely the exponents for such simplified DE-models, however the results suggest that the critical properties should be those of conventional Heisenberg ferromagnets with short-range interactions[6,9]. Further computational studies on DE-systems show that the transition may become discontinuous depending on doping level and competition with antiferromagnetic superexchange[10] similar as in the polaron model mentioned above. However, the relevance of these models for the CMR-manganites is an open problem. As pointed out above, it is possible that coupling of the magnetic subsystem to other modes may lead to a composite order-parameter and different critical properties of the PM-FM transition. Experimental evidence for ferromagnetic clusters above the Curie temperature $T_C$[11], and the observations of inhomogeneity and phase separation[12] suggest that ferromagnetic long-range order may be established by percolation of ferromagnetic regions when lowering temperature[2]. Such magnetic inhomogeneities in the spin system may result in reduced local effective topological dimensionality[13] leading also to different critical behavior. Hence, the critical properties of the paramagnetic-ferromagnetic phase transitions in manganites pose one of the important fundamental problems. On the other hand, the vast variability of competing mechanisms, which may influence the magnetic ordering, may also yield other types of PM-FM transitions for different systems (i.e. different cations at the rare-earth site) in this class of materials.

On experimental front, studies[14-25] of critical behavior of manganites near the paramagnetic-ferromagnetic phase transition using a variety of techniques have yielded a wide range of values for the magnetization critical exponent $\beta$. The values range from about 0.3 to 0.5, which embraces mean-field ($\beta$ = 0.5), 3D isotropic nearest-neighbor Heisenberg ($\beta$ = 0.365), and 3D Ising ($\beta$ = 0.325) estimates. Static dc-magnetization measurements[20-25], in addition to $\beta$, also yield critical exponents $\gamma$ and $\delta$ for initial susceptibility $\chi$ (T) and the critical isotherm M ($T_C$, H), respectively. However, they failed to determine a unique universality class for the phase transition of these manganites. Very low values of $\beta$ (0.095 for $LaMnO_{3.14}$[23], and 0.14 ± 0.02 for $La_{0.7}Ca_{0.3}MnO_3$[24]) obtained from the static magnetization measurements suggested that the paramagnetic-ferromagnetic transition in these compounds is of first order. Further, a first-order PM-FM phase transition has been reported[26] for $La_{0.7}Ca_{0.3}MnO_3$ based on the sign of the slope of the isotherm plots H/M vs. $M^2$. Interestingly, a continuous transition has been reported[25] for $La_{0.8}Ca_{0.2}MnO_3$. Recently, a tricritical point has been identified[27] in $La_{1-x}Ca_xMnO_3$ phase diagram at x=0.4, thus marking a boundary between first-order and second order phase transitions. Here we present a precise estimation of critical exponents of a $Nd_{0.6}Pb_{0.4}MnO_3$ single crystal from magnetization data in the asymptotic critical region and show that the critical exponents are 3D short-range Heisenberg like in this particular compound. The result suggests that the critical properties of manganites may vary from such a conventional behavior in compounds where a simple short-ranged ferromagnetic coupling rules the behavior to unconventional or first order transitions in compounds with several competing interactions.

## *II. EXPERIMENTAL*

Single crystals needed for the measurement were grown by flux method employing $PbO/PbF_2$ solvent[28]. Crystals with various lead concentrations were grown. Compositional analysis was carried out by Inductively Coupled Plasma Atomic Emission Spectroscopy (ICPAES). Phase purity was confirmed by x-ray diffraction. Among these crystals $Nd_{0.6}Pb_{0.4}MnO_3$, which is in ferromagnetic-metallic regime, was chosen for present investigation. This compound crystallizes in cubic structure with space group $Pm\bar{3}m$ and lattice constant a = 7.736(1) Å. Colossal magnetoresistance, defined as [R(H)-R(0)]/R(0), is found to be –86% under a field of 5 T near Curie temperature in this compound. These properties demonstrate that the chosen system belongs to the mixed-valent manganites with CMR. Single crystal structure determination has revealed that the MnO$_6$ octahedra are nearly perfect with the Mn$_1$-O$_1$ bond length = 1.934 Å, Mn$_1$-O$_2$ bond length = 1.938 Å and O$_1$-Mn$_1$-O$_2$ bond angle = 179°[29]. These lattice properties indicate that distortions in the Mn-O-Mn sub-lattice away from the ideal perovskite-lattice are weak. Extensive magnetization measurements $M(T,H)$ were performed on a well characterized Nd$_{0.6}$Pb$_{0.4}$MnO$_3$ single crystal of dimension 2×1.65×0.75 mm$^3$ in external static magnetic fields H up to 48 kOe in the temperature range 147 K ≤ T ≤ 167 K encompassing the critical region near the paramagnetic-ferromagnetic phase transition using a SQUID magnetometer. The data were collected in temperature steps of 0.5 K with the magnetic field direction along the length of the crystals.

## III. RESULTS AND DISCUSSIONS

According to the scaling hypothesis, the second-order phase transition around the Curie point $T_C$ is characterized by a set of interrelated critical exponents, $\alpha$, $\beta$, $\gamma$, and $\delta$ etc., and a magnetic equation of state[30]. The temperature dependence of the spontaneous magnetization $M_S(T) = \lim_{H \to 0} M$ just below $T_C$ is governed by exponent $\beta$ through the relation,

$$M_S(T) = M_0(-\varepsilon)^\beta, \quad \varepsilon < 0, \tag{1}$$

and that of the inverse initial susceptibility $\chi_0^{-1}(T) = \lim_{H \to 0}(H/M)$ just above $T_C$ by $\gamma$ through,

$$\chi_0^{-1}(T) = (h_0/M_0)\,\varepsilon^\gamma, \quad \varepsilon > 0 \tag{2}$$

and the exponent $\delta$ relates $M$ and $H$ at $T_C$ as

$$M = D\,H^{1/\delta}, \quad \varepsilon = 0, \tag{3}$$

where $\varepsilon = (T - T_C)/T_C$ and $M_0$, $h_0/M_0$ and $D$ are the critical amplitudes.

The magnetic equation of state in the critical region is written as

$$M(H, \varepsilon) = \varepsilon^\beta f_\pm(H/\varepsilon^{\beta+\gamma}), \tag{4}$$

where $f_+$ for $T > T_C$ and $f_-$ for $T < T_C$, respectively, are regular functions. Equation (4) implies that $M/\varepsilon^\beta$ as a function of $H/\varepsilon^{\beta+\gamma}$ falls on two universal curves, one for temperatures above $T_C$ and the other for temperatures below $T_C$.

Fig. 1 shows the Arrott plot, $M^2$ vs $H/M$, constructed from the raw M-H isotherms after correcting the external magnetic field for demagnetization effects. According to mean-field theory, near $T_C$ these curves should show a series of straight lines for different temperatures and the line at $T = T_C$ should pass through the origin. In the present case the curves in the Arrott plot are nonlinear indicating that the mean-field theory is invalid. Therefore, the values of $M_S(T,0)$ and $\chi_0^{-1}(T,0)$ were determined using a modified Arrott plot[31], in which $M^{1/\beta'}$ is plotted versus $(H/M)^{1/\gamma'}$ (Fig 2). As trial values, we have chosen $\beta' = 0.365$ and $\gamma' = 1.336$, the critical exponents of the Heisenberg model. As this plot results in nearly straight lines (for sufficiently high fields) a linear extrapolation from fields above 2 kOe to the intercepts with the axes $M^{1/\beta'}$ and $(H/M)^{1/\gamma'}$ gives the values of spontaneous magnetization $M_S(T,0)$ and inverse susceptibility $\chi_0^{-1}(T,0)$, respectively. These values as functions of temperature are plotted in Fig. 3. The continuous curves in Fig. 3 show the power law fits according to eqs. (1) and (2) to $M_S(T, 0)$ and $\chi_0^{-1}(T)$, respectively. This gives the values of $\beta = 0.369 \pm 0.02$ with $T_C = 156.4 \pm 0.2$ K (eq. (1)) and $\gamma = 1.334 \pm 0.06$ with $T_C = 156.15 \pm 0.3$ (eq. (2)). These results show that our trial values are very close to the correct critical exponents. Moreover, the curve passing the origin in Fig. 2 belongs to $T = 156.5$ K, which is near to the $T_C$ values obtained from the fits corresponding to Fig. 3. Thus, the results give support that the 3D Heisenberg universality class governs the phase transition in $Nd_{0.6}Pb_{0.4}MnO_3$.

Alternatively, the values of $T_C$, $\beta$ and $\gamma$ are also obtained by the Kouvel-Fisher (KF) method[32]. According to this method, plots of $M_S(dM_S/dT)^{-1}$ vs. $T$ and $\chi_0^{-1}(d\chi_0^{-1}/dT)^{-1}$ vs. $T$ should yield straight lines with slopes $1/\beta$ and $1/\gamma$, respectively. When extrapolated to the ordinate equal to zero, these straight lines should give intercepts on their $T$ axes equal to the Curie temperature. In Fig. 4 these plots are shown. The straight lines obtained from a least-square fit to the data give the values of $\beta = 0.374 \pm 0.006$, $T_C = 156.47 \pm 0.06$ and $\gamma = 1.329 \pm 0.003$, $T_C = 156.16 \pm 0.06$, respectively. In order to check whether the present composition is close to a tricritical point in $Nd_{1-x}Pb_xMnO_3$ phase diagram, the exponent values $\beta=0.25$ and $\gamma=1$ have been tried by fixing these values in fitting functions. These yield a very poor fit to the data. Further, specific heat measurements[33] on $Nd_{1-x}Pb_xMnO_3$ (x=0.2, 0.3, 0.4 and 0.5) did not show any signature of a first order transition. In fact, broad cusps of the specific heat observed at the respective ferromagnetic transition temperatures are consistent with Heisenberg-like behavior with negative critical exponent for the specific heat ($\alpha < 0$). This indicates that although $Nd_{0.6}Pb_{0.4}MnO_3$ has a rather low $T_C$, it is distinct from $La_{0.6}Ca_{0.4}MnO_3$.

The value of $\delta$ has been directly obtained by plotting the critical isotherm. Fig. 5 shows $M_S(156.5 K, H)$ vs. $H$ on a log-log scale. According to eq. (3), this should be a straight line with slope $1/\delta$. From the linear fit we obtained $\delta = 4.547 \pm 0.1$. The critical exponents have to fulfill the Widom scaling relation[34]

$$\delta = 1 + \gamma/\beta. \tag{5}$$

Using the above determined values of $\beta$ and $\gamma$, eq. (5) yields $\delta \approx 4.61 \pm 0.3$ for $\gamma$, $\beta$ evaluated according Fig. 3 and $\delta \approx 4.55 \pm 0.1$ for $\gamma$ and $\beta$ obtained by the KF method. Thus, the critical exponents found in this study obey the Widom scaling relation remarkably well.

In order to check whether our data in the critical region obey the magnetic equation of state (4), $M/\varepsilon^{\beta}$ as a function of $H/\varepsilon^{\beta+\gamma}$ is plotted in Fig. 6 using the values of critical exponents and $T_C$ obtained from above analysis. The inset shows the same results on a log-log plot. It can be clearly seen that all the points fall on two curves, one for $T < T_C$ and other one is for $T > T_C$. This corroborates that the obtained values of the critical exponents and $T_C$ are reliable and in agreement with the scaling hypothesis.

The values of critical exponents of $Nd_{0.6}Pb_{0.4}MnO_3$ (present work), the conventional ferromagnet Ni[35], theoretical values obtained for different models[35], and values obtained from the static magnetization measurements reported in literature[20-25] for other manganites are listed in Table 1 for comparison. Clearly, the values of critical exponents found for $Nd_{0.6}Pb_{0.4}MnO_3$ completely agree with those of conventional Heisenberg ferromagnets. Secondary effects on the PM-FM transition due to magnetic anisotropies or dipolar long-range couplings are expected to be small or unobservable, as in similar systems[35]. In particular, a crossover to another universality class due to magnetic anisotropy should be unobservable in this virtually cubic system[36]. The lower cut-off used for $\varepsilon$ avoids any such problems. According to our analysis this particular manganite system displays consistent critical properties for the paramagnetic-ferromagnetic transition placing it into the expected universality class of isotropic 3D ferromagnets. This result is not trivial: the critical exponents are governed by lattice dimension (D=3 in present case), dimension of order parameter (n=3, magnetization) and range of interaction (short-range, long-range, or infinite)[37,38]. Thus, our results show that the 'additional degrees of freedom' related to the fact that the ferromagnetic coupling is mediated by itinerant charge carriers do not mimic long-range interactions. The $\beta$ value reported for a $La_{0.7}Sr_{0.3}MnO_3$ single crystal[20] is close to that of a conventional ferromagnet and consistent with the Heisenberg model but $\gamma$ in Ref. 20 is not close to any theoretical model shown in Table 1. On the other hand, mean-field-theory values are obtained[21] for polycrystalline $La_{0.8}Sr_{0.2}MnO_3$. A first-order transition in $La_{0.7}Ca_{0.3}MnO_3$[26] and a cross-over to a continuous phase transition on either side of the phase diagram[25,27] suggest that Ca-doped manganites are distinct from other manganites. Some theoretical models[10,39] predict that such a cross-over from continuous to first-order transitions may occur. However, in a recent study[40] on a $La_{0.7}Ca_{0.3}MnO_3$ single crystal, it is proposed that the transition should be viewed in the context of Griffiths singularity, which arises when disorder suppresses a magnetic transition. Here we would like to remark on critical behavior of other half-metallic oxides like $CrO_2$[41] and $Sr_2FeMoO_6$[42]. It has been found that both materials belong to the 3D Heisenberg ferromagnet universality class. Our study demonstrates that a mixed-valent half-metallic ferromagnetic manganite also shows these conventional critical properties. This finding should provide a point of reference also for an understanding of the *anomalous* behavior in some of the manganites with unconventional PM-FM transitions. Hence, this result is useful to identify the further effects, like polaronic coupling of charge-carriers to other modes and disorder, which may be necessary to explain the various anomalies and the CMR in this class of ferromagnetic manganites.

From the experimental studies on manganites reported in literature so far, it remains unclear whether continuous paramagnetic-ferromagnetic transitions, if present, should generally belong to the universality class of the Heisenberg model with short-range couplings. The wide disparity of critical properties reported in literature for manganites calls for more experimental studies on high purity samples, in particular single crystal investigations with different compositions.

*IV. CONCLUSIONS*

In summary, we have studied the critical behavior of $Nd_{0.6}Pb_{0.4}MnO_3$ in the temperature region around $T_C$ from dc magnetization measurements and determined the values of $T_C$, $\beta$, $\gamma$ and $\delta$, which provide a consistent description of the continuous paramagnetic-ferromagnetic transition according to scaling laws. The values of the critical exponents are very close to the values for the universality class of 3D Heisenberg ferromagnets with short-range interaction.

## V. ACKNOWLEDGEMENTS

M. S. thanks Alexander von Humboldt foundation for a fellowship. N.G is grateful to the financial support of CSIR extramural research grant.

**Table 1**. Comparison of critical parameters of $Nd_{0.6}Pb_{0.4}MnO_3$ with the conventional ferromagnet Ni, different theoretical models, and manganites reported in literature. Abbreviations: SC, single crystal; PC, polycrystalline, NS, not specified; NEI, not estimated independently.

| Material | Ref. | $|\varepsilon|$ range for fit | $T_C$ (K) | β | γ | δ |
|---|---|---|---|---|---|---|
| $Nd_{0.6}Pb_{0.4}MnO_3$, SC | This Work | 0.006 – 0.05 | 156.47 ± 0.06 | 0.374 ± 0.006 | 1.329 ± 0.003 | 4.547 ± 0.1 |
| Ni, PC | [35] | 0.00024 - 0.012 | 635.53 ± 0.02 | 0.395 ± 0.01 | 1.345 ± 0.01 | 4.35 ± 0.06 |
| Mean-field theory | [35] | | | 0.5 | 1.0 | 3.0 |
| 3D Ising model | [35] | | | 0.325 | 1.24 | 4.82 |
| 3D Heisenberg model | [35] | | | 0.365 | 1.336 | 4.80 |
| Tricritical mean-field theory | [27] | | | 0.25 | 1 | 5 |
| $La_{0.7}Sr_{0.3}MnO_3$, SC | [20] | 0.002-0.03 | 354.0 ± 0.2 | 0.37 ± 0.04 | 1.22 ± 0.03 | 4.25 ± 0.2 |
| $La_{0.8}Sr_{0.2}MnO_3$, PC | [16] | NS | 315.74 | 0.50 ± 0.02 | 1.08 ± 0.03 | 3.13 ± 0.2 |
| $La_{0.67}Ba_{0.33}MnO_3$, PC | [21] | NS | 338.1 ± 0.2 | 0.464 ± 0.003 | 1.29 ± 0.02 | NEI |
| $La_{0.8}Ca_{0.2}MnO_3$, SC | [25] | NS | NS | 0.36 | 1.45 | NEI |
| $LaMnO_{3.14}$, PC | [23] | NS | 141 | 0.095 | 1.47 | NEI |
| $La_{0.7}Ca_{0.3}MnO_3$, SC | [24] | NS | 222 ± 0.2 | 0.14 ± 0.02 | 0.81 ± 0.03 | 1.22 ± 0.02 |
| $La_{0.6}Ca_{0.4}MnO_3$, PC | [27] | 0.002 – 0.04 | 265.5 ± 0.5 | 0.25 ± 0.03 | 1.03 ± 0.05 | 5.0 ± 0.8 |

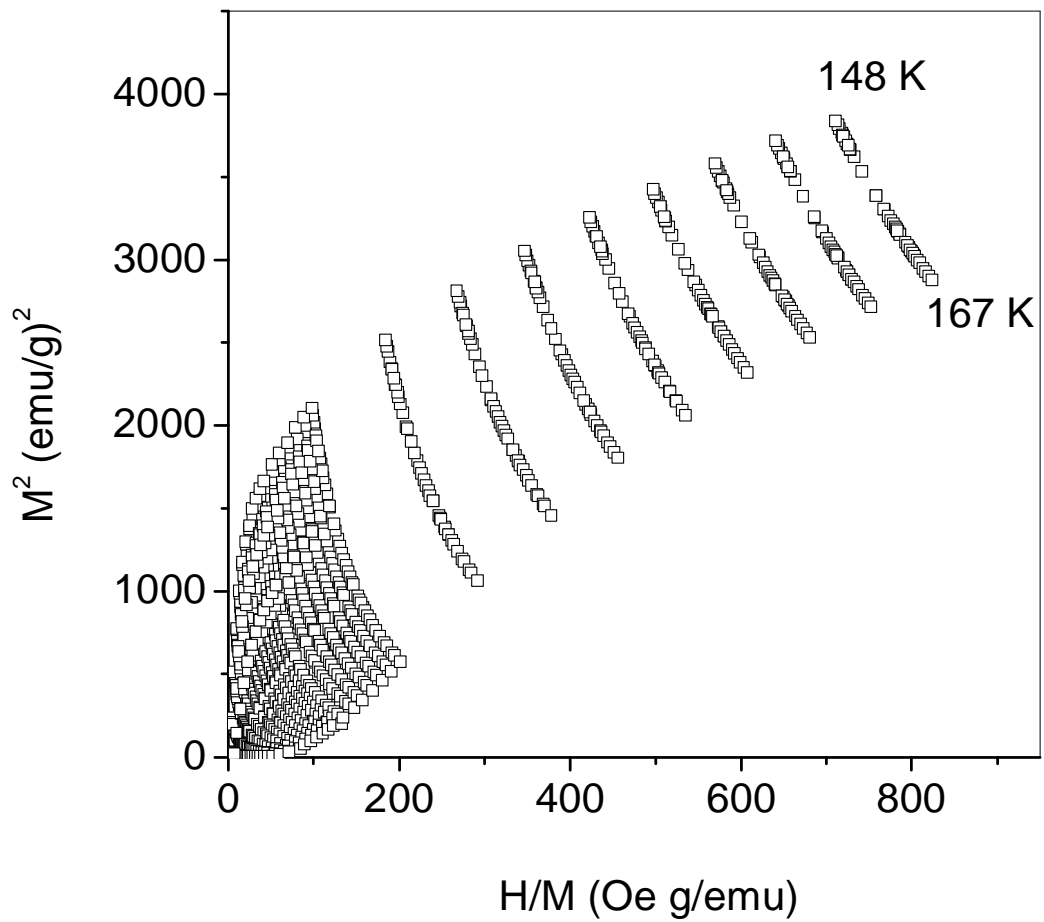

Fig. 1: Isotherms of $M^2$ vs. $H/M$ of $Nd_{0.6}Pb_{0.4}MnO_3$ at different temperatures close to the Curie temperature ($T_C = 156.5$ K).

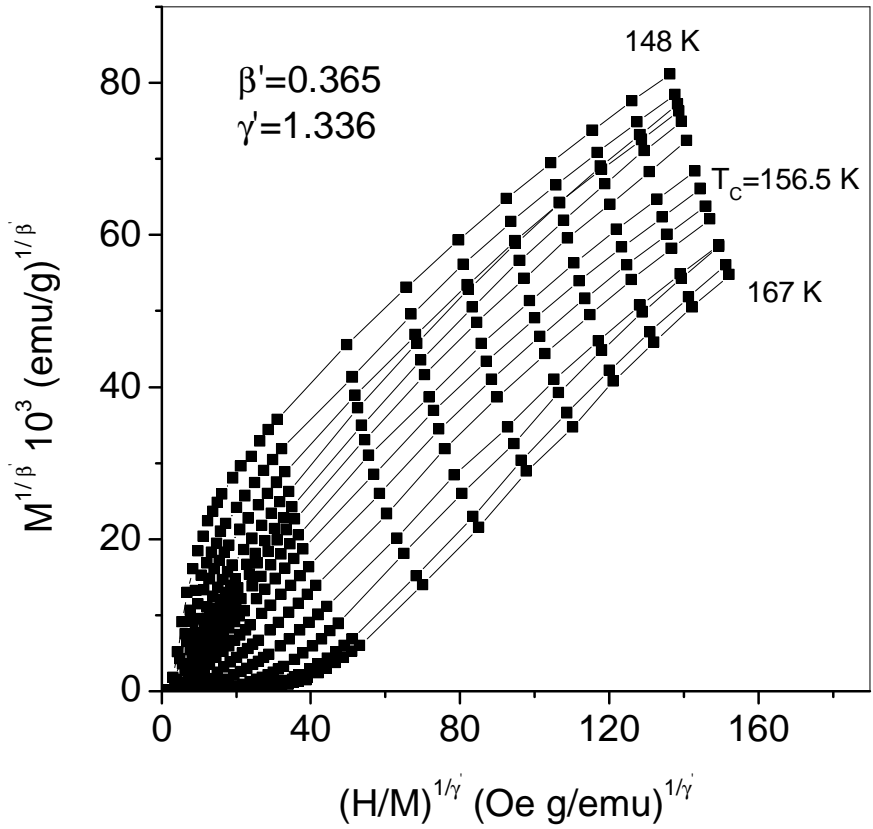

Fig .2 Modified Arrott plot isotherms $M^{1/\beta'}$ vs. $(H/M)^{1/\gamma'}$, with $\beta'$ =0.365 and $\gamma'$ =1.336. Some of the isotherms are omitted in this figure for clarity. T ~ $T_C$ = 156.5 K is the value obtained in this study.

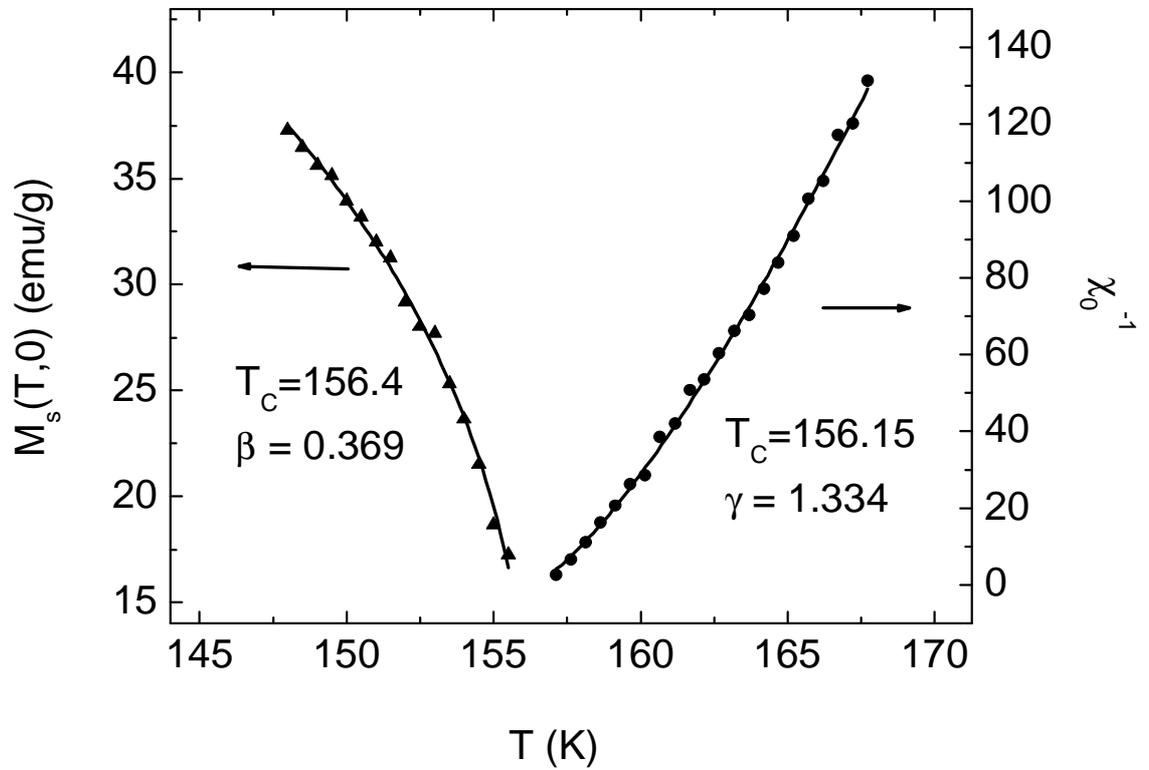

Fig. 3 Temperature variation of the spontaneous magnetization $M_S$ (triangle) and the inverse initial susceptibility $\chi_0^{-1}$ (circles) along with fits obtained for power laws.

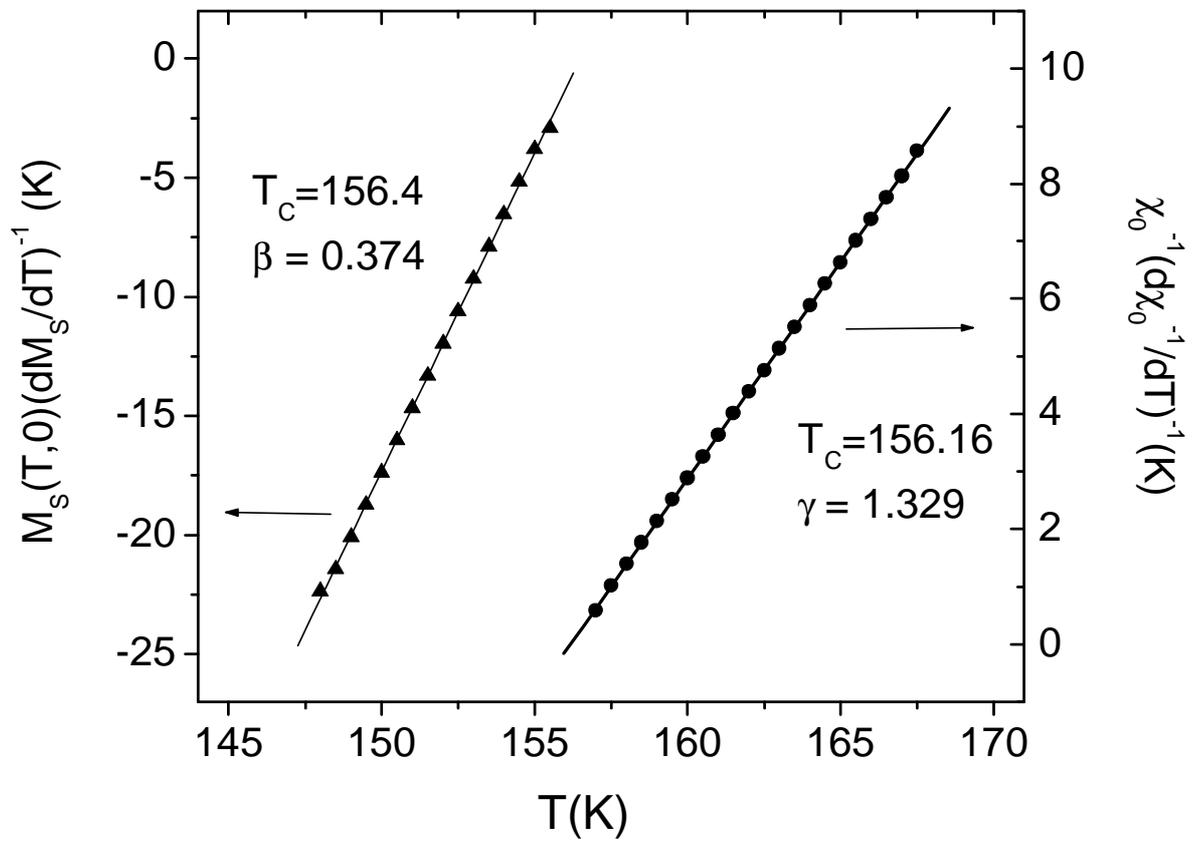

Fig. 4 Kouvel-Fisher plot for the spontaneous magnetization $M_S$ (triangles) and the inverse initial susceptibility $\chi_0^{-1}$ (circles).

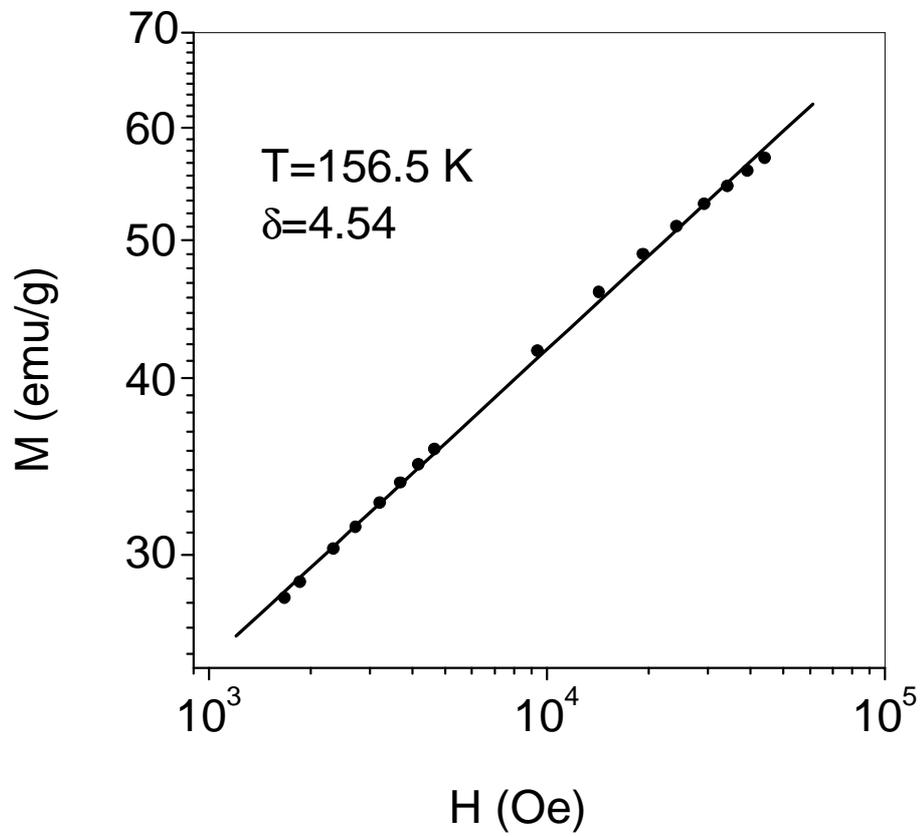

Fig. 5 *M* vs. *H* on a log-log scale at 156. 5 K, i.e, T ~ $T_C$. The straight line is the linear fit following eq. (3).

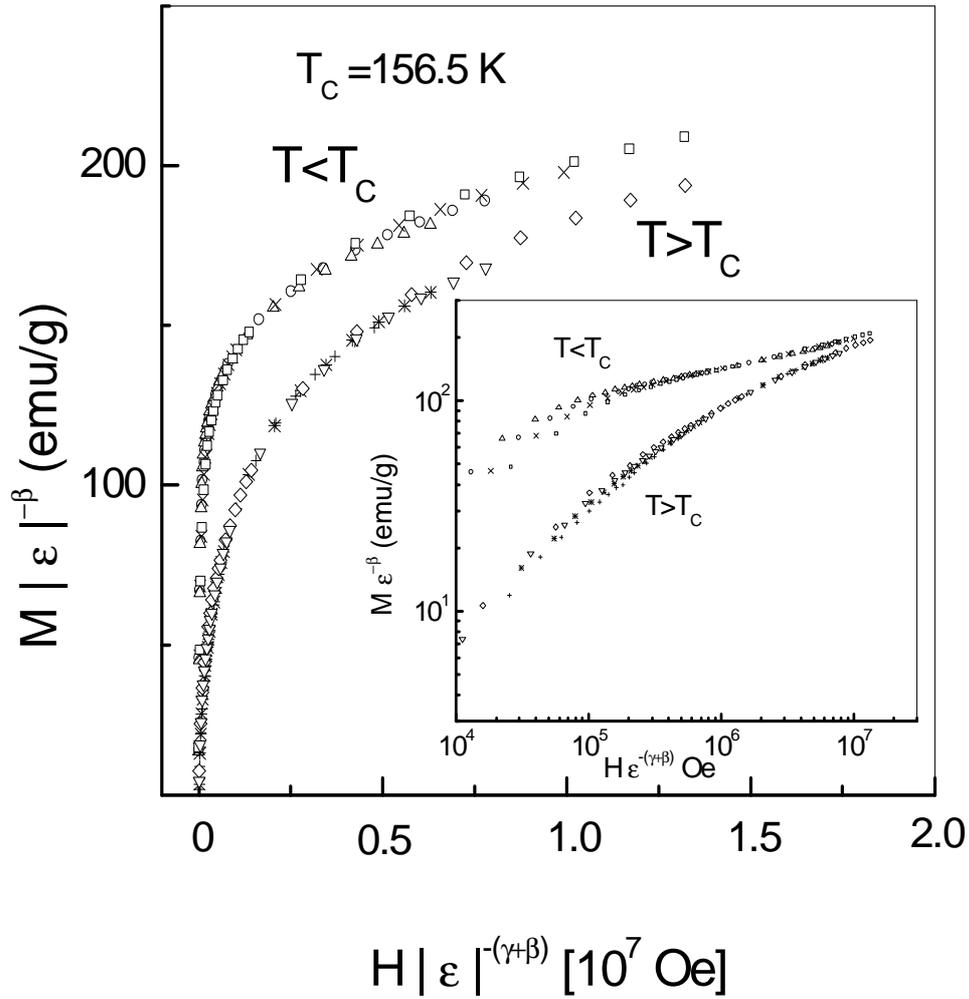

Fig. 6 Normalized isotherms of $Nd_{0.6}Pb_{0.4}MnO_3$ below and above Curie temperature ($T_C$ = 156.5 K) using $\beta$ and $\gamma$ determined as explained in the text. Inset shows the same plot on a log-log scale. $|\varepsilon| = |T - T_C|/T_C$. Different symbols on the same scaling curve $f_+$ ($f_-$) corresponds to different temperatures; 148 K (up triangles), 149 K (circles), 150 K (×), 151 K (squares) and 162 K (diamonds), 164 K (down triangles), 165 K (*), 165.5 K (+).